\documentclass[prd,twocolumn]{revtex4}
\usepackage{bm}
\usepackage{epsfig}
\usepackage{amsmath}
\newcommand{\be}{\begin{equation}}
\newcommand{\ee}{\end{equation}}
\newcommand{\bml}{\begin{mathletters}}
\newcommand{\eml}{\end{mathletters}}
\newcommand{\bes}{\begin{subequations}}
\newcommand{\ees}{\end{subequations}}

\newcommand{\bi}{\begin{itemize}}
\newcommand{\ei}{\end{itemize}}

\begin{document}
\title{Leptogenesis in a model of Dark Energy and Dark Matter}
\author{P. Q. Hung}
\email[]{pqh@virginia.edu}
\affiliation{Dept. of Physics, University of Virginia, \\
382 McCormick Road, P. O. Box 400714, Charlottesville, Virginia 22904-4714, 
USA}
\date{\today}
\begin{abstract}
A recent model of dark energy and dark matter was proposed,
involving a new gauge group $SU(2)_Z$ whose
coupling grows strong at a scale $\Lambda_Z \sim 10^{-3}\,eV$, a result
which is obtained from a simple assumption that its initial value
at some high energy scale $M \sim 10^{16}\,GeV$ is of 
the order of a typical Standard Model (SM) coupling at a similar scale.
(This assumption comes naturally from an embedding of $SU(2)_Z$ and the
SM into a grand unified group $E_6$.)
It is found that the proposed model contains a SM lepton-number violating
Yukawa coupling involving a scalar ``messenger field''
${\tilde{\bm{\varphi}}}^{(Z)}$ (which carries both $SU(2)_Z$ and
electroweak quantum numbers), a $SU(2)_Z$
fermion $\psi^{(Z)}$ and a SM lepton $l$. 
The interference between the tree-level and one-loop decay
amplitude for ${\tilde{\bm{\varphi}}}^{(Z)} \rightarrow \psi^{(Z)} + l$ 
generates
a SM lepton asymmetry which is subsequently converted into a
baryon asymmetry through electroweak sphaleron processes. It turns out
that a non-vanishing lepton asymmetry is linked to the breaking
of a ``custodial'' symmetry in the shadow fermion sector, namely
a symmetry between $\psi^{(Z)}_{1}$ and $\psi^{(Z)}_{2}$.
Furthermore, the mass of the messenger field can be less than a few TeV's,
making it accessible to searches at future colliders: the ``progenitor''
of a net lepton number (and hence a net baryon number) could
possibly be found and identified experimentally.


\end{abstract}
\maketitle

\section{Introduction}

The origin of the baryon asymmetry $\eta_{B} =
(n_{B}-n_{\bar{B}})/n_{\gamma}
=6.1 \pm 0.3 \times 10^{-10}$ is one of the most puzzling questions
in Cosmology. A universe which was initially baryon-antibaryon 
symmetric will leave a baryon number of at least eight
orders of magnitude smaller than the previous value.
A set of criteria which must be satisfied by any model
of baryogenesis was laid out by Sakharov \cite{sakharov}
almost forty years ago for the purpose of calculating this
asymmetry. Grand Unified Theories (GUT) contain the necessary ingredients
for baryogenesis \cite{gut}: the out-of-equilibrium decay
of a massive particle which violates baryon number as well as CP.
However, there are several issues with this scenario. The most
serious one is the presence of 
electroweak (EW) sphaleron processes 
at temperatures greater than the electroweak scale which
conserve $B-L$ but violate $B+L$, where $B$ and $L$ are the
baryon and lepton number respectively. It implies that
any $B+L$ asymmetry generated by GUT mechanisms would be
``washed out'' by the EW sphaleron processes \cite{kuzmin}.
It was then realized that one might need $B-L$ to be violated
itself in order to generate any baryon asymmetry.

What might be the possible sources of $B-L$ violation?

A very promising mechanism under the name of {\em leptogenesis}
was proposed in which an out-of-equilibrium decay of
a heavy Majorana neutrino which violates $B-L$ is
responsible for a lepton asymmetry \cite{fukugita}. If this 
happens at high enough temperatures while the EW sphaleron processes 
are still in equilibrium, this lepton asymmetry can be converted
into a baryon asymmetry. In these scenarios, the lepton number is
associated with Standard Model (SM) leptons and the baryon
number is associated with SM quarks. Let us recall
that, in the SM, $B$ and $L$ are violated because the
SM baryonic current, $J_{\mu}^{B}$ and SM leptonic current,
$J_{\mu}^{L}$, have an anomaly given by
\be
\label{anomaly}
\partial^{\mu} J_{\mu}^{B} = \partial^{\mu} J_{\mu}^{L}=
(\frac{n_f}{32\,\pi^2})\bigglb( \, -g^{2}\, W_{\mu \nu}^{a} 
\tilde{W}^{a\,\mu \nu}+ g^{'2} B_{\mu \nu} 
\tilde{B}^{\,\mu \nu} \biggrb) \,,
\ee
where $W_{\mu}^{a}$ and $B_{\mu}$ are the $SU(2)_L$ and $U(1)_Y$ gauge
bosons respectively.

The aforementioned leptogenesis scenarios have spawned a considerable
amount of very interesting works, especially in connection with
constraints on neutrino masses (see e.g. the excellent review
by Buchm\"{u}ller, Peccei, and Yaganida in \cite{fukugita}). It goes without
saying that much remains to be done along this path. From an
experimental point of view, the question of whether neutrinos
are Majorana or Dirac is far from being settled, with more
experiments being planned to study this issue.
The attractive and popular see-saw mechanism
which gives rise to small neutrino masses, contains Majorana 
neutrinos, with the heavier ones being candidates for the
leptogenesis scenario. (There are scenarios in which heavy
Dirac neutrinos could be responsible for leptogenesis
\cite{lindner}.) In view of these issues, it might
be interesting to investigate alternative scenarios
of leptogenesis. Could there be a mechanism of leptogenesis in which
the $B-L$ violation comes from the decay of some particle
other than the heavy Majorana neutrino? After all, it is
the SM lepton number violation which is at the heart of the
matter, no matter what its source might be. 
Can one test this new scenario in
terms of its particle physics implications?

There is indeed such a particle as described in \cite{hung2}.
It arises in the construction of a model of dark energy and
dark matter \cite{su2}, \cite{hung2}. We summarize below
the essence of that model in order to motivate the model
of leptogenesis presented in this paper.

In recent papers \cite{su2}, \cite{hung2}, a Quintessence model 
was proposed in which the quintessence field is an 
axion-like particle, $a_Z$, of a spontaneously broken
global $U(1)_{A}^{(Z)}$ symmetry whose potential is induced by the
instantons of a new unbroken gauge group $SU(2)_Z$. The $SU(2)_Z$ coupling
becomes large at a scale $\Lambda_Z \sim 10^{-3}\,eV$ starting
from an initial value $M$ at high energy which is of the order of the
Standard Model (SM) couplings at the same scale $M$.
This last fact could come from the following Grand Unified path
$E_6 \rightarrow SU(2)_Z \otimes SU(6)$ with $SU(6)$ ultimately
breaking down to the Standard Model, the details of which is
given in \cite{hung3}.
The scenario which was proposed in \cite{su2}, \cite{hung2} is one
in which $a_Z$ gets trapped in a false vacuum of an instanton-induced
potential with a vacuum energy density $\sim (10^{-3}\,eV)^4$. This model
of quintessence mimics a universe which is dominated by a cosmological
constant and cold dark matter. In fact, the most recent analyses
from the Supernova Legacy Survey (SNLS) and WMAP \cite{snls}, \cite{WMAP}
fits a flat $\Lambda\,CDM$ with a constant equation of state
$w = -0.97 \begin{array}{c}
+0.07 \\ - 0.09 \end{array}$. As noticed in \cite{WMAP}, even without
the prior that the universe is flat, the combination of WMAP,
large scale structure and supernova data gives $w= -1.06 \begin{array}{c}
+ 0.13 \\ -0.08 \end{array}$

As discussed in \cite{hung2}, our model, beside providing a scenario
for the dark energy, contains several other phenomenological
and cosmological consequences, two of which involve a candidate
for the cold dark matter and a candidate for a new scenario of
leptogenesis. The purpose of the present paper is to present a
detailed description of this new mechanism of leptogenesis.

These aforementioned candidates depend on each other in an interesting way.
The $SU(2)_Z$ fermions (the shadow fermions), which transform as $(3,1,0)$ under
$SU(2)_Z \otimes SU(2)_L \otimes U(1)_Y$,
would not have any interaction with the SM particles (the
visible sector) (other
than the gravitational one) if it were not for the presence
of a messenger scalar field $\tilde{\varphi}^{(Z)} =(3,2,-1/2)$
in our model. As discussed in \cite{hung2}, this presence
manifests itself in a variety of ways: it helps maintain thermal 
equilibrium between the $SU(2)_Z$ and SM plasmas (so that
the two sectors possess a common temperature) until it
drops out of thermal equilibrium. Its decay into a SM lepton
plus a $SU(2)_Z$ fermion, as we shall see below, generates an
asymmetry of the {\em SM lepton number} which is subsequently reprocessed
into a baryon number asymmetry through the electroweak sphaleron
process. Furthermore, it will be seen below that the asymmetry
depends on $(\frac{m_{\psi^{(Z)}_{2}}^2 -
m_{\psi^{(Z)}_{1}}^2}{m_{\tilde{\varphi}_{1}}^2})$
besides other factors such as CP phase factors, etc..., where
$m_{\psi^{(Z)}_{1,2}}$ and $m_{\tilde{\varphi}_{1}}$ are
the masses of the $SU(2)_Z$ fermions and messenger field
respectively. The non-vanishing asymmetry is seen to be linked
to the breaking of a shadow ``custodial'' symmetry $SU(2)_{shadow}$
by the difference in mass among the two shadow fermions.
By requiring the SM leptonic asymmetry to be of 
order $10^{-7}$, various upper bounds on the messenger mass
are obtained. The messenger field can be as light as several
hundreds of GeVs which makes it an interesting prospect for
a search at future colliders such as the LHC.

This scenario of leptogenesis is drastically
different from the ``standard'' one in that here it is a scalar field
whose decays violate SM lepton numbers instead of the decays
of the customary right-handed Majorana neutrinos. In some
sense, it is reminescent of the color-triplet Higgs scalar
of $SU(5)$ with the difference being that, in our case, only
SM lepton number is violated. Let us recall that in
``standard'' scenarios with two heavy particles, one being much
heavier than the decaying particle, the computation of the
asymmetry gives rise to a factor which is proportional to
$1/x$ with $x \equiv (m_{heavy}/m_{light})^2$. In these models,
SM particles which have masses much smaller than $m_{light}$,
contribute a negligible amount to the asymmetry. In contrast,
our model contains fermions, $\psi^{(Z)}_{i}$, whose
masses are not too much smaller than that of the ``light'' messenger scalar
field and whose contribution in the asymmetry turns out
to be proportional to $(m_{\psi^{(Z)}_{i}}/m_{\tilde{\varphi}_{1}})^{2}$.
This is a contribution which greatly dominates over that of
$(m_{\tilde{\varphi}_{1}}/m_{\tilde{\varphi}_{2}})^2 < 10^{-26}$ in our
model. These points will be made clear below.

We would like to mention that there exists models of
baryogenesis where there is an asymmetry between SM particles
and e.g. particles that are not affected by the electroweak sphalerons
\cite{kuzmin2} or scalar condensates \cite{dodelson}. Our model
is similar in spirit but is entirely different from the
aforementioned interesting models.

In this paper, we will lay out the groundwork for the computation
of the SM lepton asymmetry from $\tilde{\varphi}^{(Z)}$ decays. First, we
will give a brief summary of the salient points of the $SU(2)_Z$ model
(nicknamed {\em QZD}).
In particular, we will focus on the particle content and the
related interactions which are most relevant for this paper. 
We will discuss the reason for having two scalars:
$\tilde{\varphi}_{1,2}^{(Z)}$.
We then proceed with the computation, at $T=0$, of the 
SM lepton asymmetry, showing
its dependence both on the masses of the particles involved
and on the strengths of the couplings and the CP violation. 
A more complete treatment of the problem at $T \neq 0$ will
be dealt with elsewhere. Here, the main aim will be to show
that the SM lepton asymmetry, in our model, can be non-vanishing
at zero temperature.
We end with a brief discussion on a possible detection of the 
lightest scalar which is responsible for this SM lepton 
number asymmetry. An interesting feature of this model is
the fact that this ``messenger field'' which is the ``progenitor''
of the lepton asymmetry could in fact be found and identified in
future colliders such as the Large Hadron Collider (LHC).

\section{A brief review of $SU(2)_Z$}
\label{review}

In this section, we summarize the essential elements of the $SU(2)_Z$
model used in \cite{su2}, \cite{hung2}, restricting ourselves to the
non-supersymmetric case.
At some scale $M$, the gauge group is described by
\be
\label{gauge}
G_{SM} \otimes SU(2)_Z 
\ee
where $G_{SM}$ could be, for example, $SU(6)$ as in the chain
$E_6 \rightarrow SU(6) \otimes SU(2)_Z$ \cite{hung2},
\cite{hung3}, with $SU(6)$ breaking down to
$SU(3)_{C} \otimes SU(2)_L \otimes U(1)_Y$ via, e.g., the
route $SU(3)_{C} \otimes SU(3)_L \otimes U(1)$ \cite{hung2},
\cite{hung3}.
Fermion fields transform under the above gauge group as
\be
\label{fermion}
\psi_{L,R}^{SM} = (R_{L,R}, 1) \,; 
\psi_{L,R}^{(Z)} = (1,3) \,,
\ee
where $R_{L,R}$ denotes the representation of the left-handed
and right-handed SM fermions under $G_{SM}$. Notice that the
fermions of each sector are singlets under the other's gauge
group. Notice also that $SU(2)_Z$ is a vector-like
gauge group, similar to ordinary QCD. Apart from the 
obvious gravitational interactions,
the two sectors can communicate with each other through
a messenger scalar field which carries quantum numbers of
both sectors. We shall see below that one actually needs
two of such scalars, one being much heavier than the other.
In this paper, we will concentrate on the type of messenger
fields which are crucial for our leptogenesis scenario. 
They are:
\be
\label{messenger}
{\tilde{\bm{\varphi}}}_{i}^{(Z)} =({\tilde{\bm{\varphi}}}_{i}^{(Z),0},
{\tilde{\bm{\varphi}}}^{(Z),-}_{i})=
(1,2,Y_{\tilde{\varphi}}=-1,\,3) 
\ee
under $SU(3) \otimes SU(2)_L
\otimes U(1)_Y \otimes SU(2)_Z$, where $i=1,2$ and where
$Q= T_{3L} + Y/2$.
Since we wish $SU(2)_Z$ to be unbroken and to grow strong
at $\Lambda_Z \sim 10^{-3}\,eV$, we
will assume that the potential for ${\tilde{\bm{\varphi}}}_{i}^{(Z)}$ 
is such that
$\langle {\tilde{\bm{\varphi}}}_{i}^{(Z)} \rangle =0$. 
As a consequence, it will
{\em not} contribute to the breaking of the electroweak gauge
group. 
The physical masses of the messenger fields are arbitrary.
As explained in \cite{hung2}, one of the two messenger fields
is assumed to have a mass less than $1\,TeV$ so
that the $SU(2)_Z$ and SM plasmas maintain thermal
equilibrium until ${\tilde{\bm{\varphi}}}_{1}^{(Z)}$ drops
out of equilibrium and the other,
${\tilde{\bm{\varphi}}}_{2}^{(Z)}$
is assumed to be very massive, with a mass of the order of
a typical GUT scale in order for the evolution of the
$SU(2)_Z$ coupling to yield the desired features of the model.
This also turns out to be what we need for the leptogenesis scenario.

In addition to one of the above messenger fields (the heavy
one with GUT-scale mass is not included in the evolution of the coupling),
it is shown in \cite{hung2} that the following fermions are
needed in order for 
the $SU(2)_Z$ 
coupling $\alpha_Z = g_{Z}^2/4\pi$
to be of order unity at around $\Lambda_Z = 3 \times 10^{-3}\,eV$:
$\psi_{1}^{(Z)}$ and $\psi_{2}^{(Z)}$. As shown in \cite{hung2},
the masses of $\psi_{1}^{(Z)}$ and $\psi_{2}^{(Z)}$ come from
a complex scalar which is a singlet under both the SM and $SU(2)_Z$.
(As shown in \cite{hung2}, the ``axion'', which is the imaginary part
of  this complex scalar, is the quintessence field which gets trapped
in a false vacuum and yields a scenario for the dark energy.)
The vacuum expectation of the real part of that scalar is unconstrained
by present particle physics data, although a recent model of ``low scale''
inflationary scenario did put a constraint on its VEV \cite{inflation}. 
As a result, the masses of
$\psi_{1}^{(Z)}$ and $\psi_{2}^{(Z)}$ are arbitrary. However, as
it is argued in \cite{hung2}, they (or at least one of them) can
be a candidate for a WIMP cold dark matter if its mass is of
O($100-200\,GeV$). As discussed in \cite{hung2}, the most
attractive WIMP scenario in our model is one in which 
$\psi_{1}^{(Z)}$ and $\psi_{2}^{(Z)}$ are close in mass to each other,
with $m_{\psi^{(Z)}_{2}} \sim m_{\psi^{(Z)}_{1}} \sim \mathcal{O}(100\,GeV)$.
It is in this context that we will concentrate our discussion
of leptogenesis.

\begin{figure}
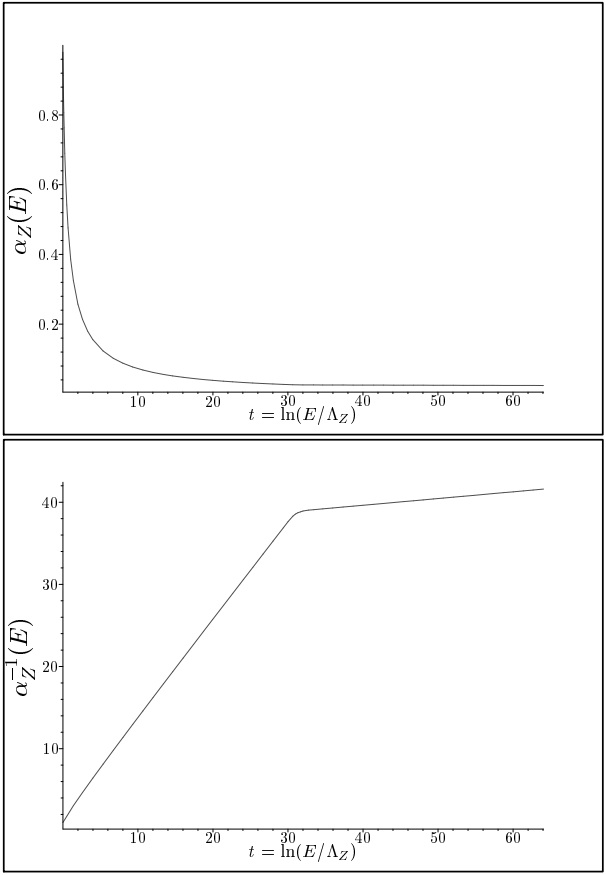

\includegraphics[angle=-90,width=8cm]{alvst_50.epsi}
\includegraphics[angle=-90,width=8cm]{alinvst_50.epsi} 
\caption{\label{fig1}$\alpha_Z(E)$ and $\alpha_Z^{-1}(E)$ versus
$t=\ln (E/\Lambda_Z)$ for $m_{\tilde{\bm{\varphi}}_{1}^{(Z)}}
= 300\,GeV$, $m_{\psi^{(Z)}_{2}} = 100\; GeV$ and
$m_{\psi^{(Z)}_{1}}= 50\;GeV$. Here $\Lambda_Z = 3 \times 10^{-3}\,eV$.}
\end{figure}

To complete this review section, we show, for illustration, a couple of graphs
of $\alpha_Z$ and $\alpha_{Z}^{-1}$ versus $E$ taken from \cite{hung2} for
a given value of ${\tilde{\bm{\varphi}}}_{1}^{(Z)}$
mass with two different values of $\psi_{1,2}^{(Z)}$ masses.
(The constraint is $\alpha_Z = 1$ at $\Lambda_Z = 3 \times 10^{-3}\,eV$.)
Although the calculations presented below are meant to be general,
we will illustrate our results with masses which are in the range
of values illustrated in Figures \ref{fig1} and \ref{fig2}.



In \cite{su2},
it was shown that the ``minimum model'' with
$\psi_{1}^{(Z)}$, $\psi_{2}^{(Z)}$ and $\tilde{\bm{\varphi}}_{1}^{(Z)}$
was sufficient as a scenario
for both the dark energy and dark matter. However, as discussed in
\cite{hung2}, a simple extension of this ``minimum model''
by including {\em one} extra heavy messenger field 
$\tilde{\bm{\varphi}}_{2}^{(Z)}$ has a far-reaching consequence:
a non-vanishing SM lepton number asymmetry as we shall see below.
We will assume that $\langle \tilde{\varphi}_{1,2}^{(Z)} \rangle =0$
and $m_{\tilde{\varphi}_2}^2 \gg m_{\tilde{\varphi}_1}^2 >0$. 
Without this extra messenger field, the asymmetry will simply vanish.
This is a well-known result of early models of baryogenesis \cite{kolb}.
It also turns out that the dominant
contributions to this asymmetry is insensitive to the value
of $m_{\tilde{\varphi}_2}$ as long as 
$m_{\tilde{\varphi}_2} \gg m_{\tilde{\varphi}_1}$. In this case,
$m_{\tilde{\varphi}_2}$ can be as large as a typical ``GUT'' scale.

Although the lepton asymmetry discussed in this paper comes
primarily from the decay of $\tilde{\bm{\varphi}}_{1}^{(Z)}$
when it drops out of thermal equilibrium, in
principle the decay of the much heavier $\tilde{\bm{\varphi}}_{2}^{(Z)}$
can also generate a SM lepton number asymmetry. However, this
asymmetry will be washed out by the inverse-decay into the
lighter $\tilde{\bm{\varphi}}_{1}^{(Z)}$ at $T> m_{\tilde{\varphi}_1}$.
This is similar to the popular leptogenesis scenario with
two Majorana neutrinos, one of which being much heavier than
the other. Because of this fact, we will focus only
on the decay of $\tilde{\bm{\varphi}}_{1}^{(Z)}$ in this paper.

We now proceed to the discussion of our model of leptogenesis.


\section{SM leptogenesis from $\tilde{\bm{\varphi}}_{1}^{(Z)}$ decays}

In this section, we will show how the introduction of two messenger
fields gives rise to the possibility of a new mechanism for SM
leptogenesis, alternative to the popular scenario in which
the SM lepton number asymmetry is generated by the decay of a 
heavy Majorana neutrino. It will be 
shown that the value of the SM lepton asymmetry
depends primarily on the ratio of the $SU(2)_Z$ fermion mass to that
of the messenger scalar field $\tilde{\varphi}_1$, namely
$(m_{i}/
m_{\tilde{\varphi}_2})^2$, where $m_{i}$
with $i=1,2$ denotes the mass of the $SU(2)_Z$ fermion. The
(by-far) subdominant contributions are found to
be proportional to $m_{\tilde{\varphi}_1}^2/m_{\tilde{\varphi}_2}^2$
and $m_{i}^2/m_{\tilde{\varphi}_2}^2$ which are less than $10^{-26}$.
Interestingly enough, as we shall see below, for
$m_{\tilde{\varphi}_2} \gg m_{\tilde{\varphi}_1}$, the asymmetry depends (beside
other factors such as the CP phases, etc..) mostly on the ratio
$(m_{i}/ m_{\tilde{\varphi}_1})^2$
and is insensitive to the exact value of $m_{\tilde{\varphi}_2}$ as long as
$m_{\tilde{\varphi}_2} \gg m_{\tilde{\varphi}_1}$. 
Let us remind ourselves that $\psi_{i}^{(Z)}$ with
$m_i = O(100\,GeV)$ could be WIMP candidates as discussed in \cite{hung2}
and $\tilde{\varphi}_1$ with a mass not-too-different from the electroweak scale
can be searched for at colliders such as the LHC. One cannot fail
but notice the interesting connection between the aforementioned ratio
which appears in the SM lepton asymmetry and the ``detectability'' of
$\psi_{i}^{(Z)}$ and $\tilde{\varphi}_{1}^{(Z)}$. We will come back
to this connection below.

Before writing down the interaction Lagrangian, let us notice a few facts.
(a) $SU(2)$ representations are real: both $3 \times 3$ and $3 \times 3^{*}$
contain a singlet. (b) $\psi^{(Z),c}_{i,L}$ transforms like a right-handed
spinor. In order to construct the diagrams shown in Fig. (\ref{fig3}), one
can write the $SU(2)_L \otimes U(1)_Y \otimes SU(2)_Z$ invariant Lagrangian
taking into account points (a) and (b) as follows:
\be
\label{yuk}
{\cal L}_{yuk}=  \sum_{i,m}( g_{\tilde{\varphi}_{1}\,m}^{(i)}\,
\bar{l}_{L}^{m}\,
\tilde{\bm{\varphi}}_{1}^{(Z)}\,\psi^{(Z)}_{i}+ 
g_{\tilde{\varphi}_{2}\,m}^{(i)}\,\bar{l}_{L}^{m}\,
\tilde{\bm{\varphi}}_{2}^{(Z)}\,\psi^{(Z)}_{i})+ H.c. \,,
\ee
where
\be
\label{psi}
\psi^{(Z)}_{i} \equiv \psi^{(Z)}_{i,R} + \psi^{(Z),c}_{i,L} \,,
\ee
and where, in general, the couplings $g_{\tilde{\varphi}_{(1,2)}\,m}^{i}$ 
are complex and where $m=1,2,3$ and $i=1,2$ refer to the 
lepton family number and the two $SU(2)_Z$ fermions respectively. 
In general,
one has twelve complex Yukawa couplings in total. We write them as
\be
\label{phase1}
g_{\tilde{\varphi}_{1}\,m}^{(1,2)}=|g_{\tilde{\varphi}_{1}\,m}^{(1,2)}|
\exp (i \alpha_{(1,2),m}) \,,
\ee 
\be
\label{phase2}
g_{\tilde{\varphi}_{2}\,m}^{(1,2)}=|g_{\tilde{\varphi}_{2}\,m}^{(1,2)}|
\exp (i \beta_{(1,2),m}) \,.
\ee 
Notice that the interaction (\ref{yuk}) violates ``lepton'' number in a 
general sense that it includes also the shadow fermions. It is then
natural to define the following Majorana shadow fermions
\be
\label{maj1}
N^{(Z)}_{i} = \psi^{(Z)}_{i,L} + \psi^{(Z,c)}_{i,L} \,,
\ee
\be
\label{maj2}
M^{(Z)}_{i} = \psi^{(Z)}_{i,R} + \psi^{(Z,c)}_{i,R} \,.
\ee
The interesting issue of Majorana shadow fermions deserves a separate
investigation. For the present purpose, we will investigate
the problem using the above Lagrangian (\ref{yuk}). With a Majorana shadow
fermion, the diagrams shown in Fig. (\ref{fig3}) are very similar to
those of the ``standard'' leptogenesis scenarios with the Majorana
particle being a decay product in our model instead of 
being the decaying particle.

As we have mentioned in our review section of the $SU(2)_Z$ model,
the case where $m_{\psi^{(Z)}_{2}} \sim m_{\psi^{(Z)}_{1}} 
\sim \mathcal{O}(100\,GeV)$ is the most attractive candidate for
a WIMP scenario. In what follows, we will focus on this particular
case, although our presentation will be sufficiently general.

We wish to compute the interference between the tree-level and one-loop
contributions to the decays 
\be
\label{decay1}
\tilde{\varphi}_{1}^{(Z)} \rightarrow \bar{\psi}^{(Z)}_{1,2}+ l
\ee
\be
\label{decay2}
\tilde{\varphi}_{1}^{(Z),*} \rightarrow \psi^{(Z)}_{1,2}+ \bar{l}
\ee 
where $l$ represents a SM lepton as shown in Fig. \ref{fig3}. 
This interference will give rise to
to an asymmetry in SM lepton number. Notice, in this regard, that only 
the SM lepton number asymmetry can be converted into a baryon asymetry
through the electroweak sphaleron since 
$\psi^{(Z)}_{1,2}$ are SM singlets.

Before computing the SM lepton number asymmetry, let us make a few
remarks concerning the out-of-equilibrium decay constraint for our
scenario. Since a more detailed discussion will be presented
elsewhere, we will summarize the essential points here. The primary
condition for a departure from thermal equilibrium is the 
requirement that the decay rate $\Gamma_{\tilde{\varphi}_{1}}
\sim \alpha_{\tilde{\varphi}_{1}} m_{\tilde{\varphi}_{1}}$,
with $\alpha_{\tilde{\varphi}_{1}} = g_{\alpha_{\tilde{\varphi}_{1}}}^2/4\pi$, 
is {\em less than} the expansion rate $H = 1.66\, g_{*}^{1/2} T^2/m_{pl}$,
where $g_{*}$ is the effective number of degrees of freedom at 
temperature $T$. As with \cite{kolb}, we can define
\be
\label{K}
K \equiv (\Gamma_{\tilde{\varphi}_{1}}/2\,H)_{T=m_{\tilde{\varphi}_{1}}}
= \frac{\alpha_{\tilde{\varphi}_{1}}\,m_{pl}}{3.3\,g_{*}^{1/2}\,
m_{\tilde{\varphi}_{1}}} \,.
\ee
When $K \ll 1$, $\tilde{\varphi}_{1}$ and $\tilde{\varphi}_{1}^{*}$ are
overabundant and depart from thermal equilibrium. Since the time
when $\tilde{\varphi}_{1}$ decays is $t \sim 
\Gamma^{-1}_{\tilde{\varphi}_{1}}$ and since $T \propto 1/\sqrt{t}$, the
temperature at the time of decay is found to be (using (\ref{K}))
$T_D \sim K^{1/2}\,m_{\tilde{\varphi}_{1}}$ \cite{kolb}. For this
scenario to be effective i.e. a conversion of a SM lepton number asymmetry
coming from the decay of $\tilde{\varphi}_{1}$ into a baryon number
asymmetry through the electroweak sphaleron process, one has to make sure
that the decay occurs at a temperature greater than $T_{EW} \sim
100\,GeV$ above which the sphaleron processes are in thermal equilibrium.
From this, it follows that $K$ cannot be arbitrarily small and
has a lower bound coming from the requirement $T_D > T_{EW}$. One obtains
\be
\label{lower}
1>\,K\, > (\frac{100\,GeV}{m_{\tilde{\varphi}_{1}}})^2 \,.
\ee
For example, if $m_{\tilde{\varphi}_{1}} \sim 500\,GeV$, the
allowed range for $K$ would be $0.04<K<1$. We will come back
to (\ref{lower}) below. 

A remark is in order concerning (\ref{lower}). The asymmetry
obtained when $K<1$ is actually the largest value for a given
scenario. However, when $K>1$ but not too different
from unity, the net lepton number asymmetry
will be diluted by a factor which is approximately $1/K$ (modulo
a factor which is less than 2 when $K$ is not too large). As we
shall see below, the upper bound on the mass of the messenger
``progenitor'' field will be lowered if we allow for the possibility
of $K>1$ . Rougly speaking, the dilution factor $1/K$ has
to be compensated by an increase in the basic asymmetry
$\epsilon^{\tilde{\varphi}_{1}}_{l}$ discussed below which can
come about when the mass of the messenger field is decreased.
This will be discussed at the end of this section
and in the section on phenomenology.

Once the out-of-equilibrium condition is fulfilled, the next thing to 
do is to estimate the SM lepton number asymmetry and relate it
to the sought-after baryon asymmetry. Since the main aim of the present
manuscript is to present a scenario for the computation of the
SM lepton number asymmetry, we first present an estimate
of that quantity in order to have some ideas on what to expect of the
magnitude of the asymmetry parameter $\epsilon^{\tilde{\varphi}_{1}}_{l}$
to be computed in our model. This asymmetry will be computed at $T=0$.
Although care should be taken to include finite temperature
corrections (see e.g. \cite{giudice}), one does not expect the final
result to be too different from the zero temperature one.
When $T<m_{\tilde{\varphi}_{1}}$ and when $K<1$,
the number density of $\tilde{\varphi}_{1}$ is approximately
$n_{\tilde{\varphi}_{1}} =T^{3}/\pi^2$ (overabundance) and the entropy is
$s = (2/45)\,g_{*} \pi^2\,T^3$, with $g_{*} \sim 114$
(including $SU(2)_Z$ light degrees of freedom). 
The decay of $\tilde{\varphi}_{1}$ and $\tilde{\varphi}_{1}^{*}$
creates a SM lepton number asymmetry per unit entropy 
$n_{LSM}/s \sim 2 \times 10^{-3}\,\epsilon^{\tilde{\varphi}_{1}}_{l}$.
For the SM with three generations and one Higgs doublet, one
has $n_{B}/s \sim -0.35\,n_{LSM}/s\,\sim -10^{-3}\,\epsilon^{\tilde{\varphi}_{1}}_{l}$,
where $n_B$ is ``processed'' through the electroweak sphaleron.
Since $m_{B}/s \sim 10^{-10}$, a rough constraint on 
$\epsilon^{\tilde{\varphi}_{1}}_{l}$ is found to be
\be
\label{constraint}
\epsilon^{\tilde{\varphi}_{1}}_{l} \sim -10^{-7} \,.
\ee
We will make use of the constraint (\ref{constraint}) to gain some insights 
into the allowed ranges of masses in our model.

The central quantity to be computed in our model is the asymmetry
\be
\label{asymmetry}
\epsilon^{\tilde{\varphi}_{1}} = \frac{\Gamma_{\tilde{\varphi}_{1}\,l}-
\Gamma_{\tilde{\varphi}_{1}^{*}\,\bar{l}}}{\Gamma_{\tilde{\varphi}_{1}\,l}+
\Gamma_{\tilde{\varphi}_{1}^{*}\,\bar{l}}} \,,
\ee
where $\Gamma_{\tilde{\varphi}_{1}\,l}$ and $\Gamma_{\tilde{\varphi}_{1}^{*}\,
\bar{l}}$ contain the sums over all three flavors of SM leptons. Also, in
the numerator of (\ref{asymmetry}),
$\Gamma_{\tilde{\varphi}_{1}\,l}$ and $\Gamma_{\tilde{\varphi}_{1}^{*}\,
\bar{l}}$ are computed up to one loop and therefore contain interferences
between the tree-level and one-loop contributions. 
It is only this interference which contributes to 
$\epsilon^{\tilde{\varphi}_{1}}_{l}$.
As usual, the decay widths in the denominator of (\ref{asymmetry}) 
are kept at tree level. To be more specific, we will define the
following asymmetries corresponding to the decay of $\tilde{\varphi}_{1}$
separately into $\psi^{(Z)}_{1}$ and $\psi^{(Z)}_{2}$.
\be
\label{asymmetry1}
\epsilon^{\tilde{\varphi}_{1}}_{1} = \frac{\Gamma(\tilde{\varphi}_{1}^{(Z)} 
\rightarrow \bar{\psi}^{(Z)}_{1}+l )-
\Gamma(\tilde{\varphi}_{1}^{(Z),*} \rightarrow \psi^{(Z)}_{1}+ \bar{l})}
{\Gamma(\tilde{\varphi}_{1}^{(Z)} 
\rightarrow \bar{\psi}^{(Z)}_{1}+l )+
\Gamma(\tilde{\varphi}_{1}^{(Z),*} \rightarrow \psi^{(Z)}_{1}+ \bar{l})} \,,
\ee
\be
\label{asymmetry2}
\epsilon^{\tilde{\varphi}_{1}}_{2} = \frac{\Gamma(\tilde{\varphi}_{1}^{(Z)} 
\rightarrow \bar{\psi}^{(Z)}_{2}+l )-
\Gamma(\tilde{\varphi}_{1}^{(Z),*} \rightarrow \psi^{(Z)}_{2}+ \bar{l})}
{\Gamma(\tilde{\varphi}_{1}^{(Z)} 
\rightarrow \bar{\psi}^{(Z)}_{2}+l )+
\Gamma(\tilde{\varphi}_{1}^{(Z),*} \rightarrow \psi^{(Z)}_{2}+ \bar{l})} \,.
\ee
In (\ref{asymmetry1},\ref{asymmetry2}), a non-vanishing value for
$\epsilon^{\tilde{\varphi}_{1}}_{1,2}$ comes from the interference between
the tree-level and one-loop contributions to the decay widths. In what
follows, we will concentrate on these interference terms.

We will denote the tree-level-one-loop interference contribution
to the decay width by $\Gamma_{\tilde{\varphi}_{1},1}^{int}$.
The one-loop contribution includes both vertex and self energy corrections
as shown in Fig. (\ref{fig3}). In what follows, we will neglect the SM lepton masses
in the one-loop calculations since their contributions to the
asymmetry is tiny, of order $m_l \, m_{\psi^{(Z)}_{1,2}}/m_{\tilde{\varphi}_{2}}^2$.
First, we concentrate on the vertex
contribution.
We obtain for 
$\tilde{\varphi}_{1}^{(Z)} \rightarrow \bar{\psi}^{(Z)}_{1}+ l$
\begin{eqnarray}
\label{int1}
\Gamma_{\tilde{\varphi}_{1},V}^{int(1)}& =& 
(\sum_{l}\,g_{\tilde{\varphi}_{1}\,l}^{(1)}\,
g_{\tilde{\varphi}_{2}\,l}^{(1)*}\,
\,\sum_{m}\,g_{\tilde{\varphi}_{1}\,m}^{(1)*}\,
g_{\tilde{\varphi}_{2}\,m}^{(1)})\, I^{(1)}\nonumber \\
&&+ (\sum_{l}\,g_{\tilde{\varphi}_{1}\,l}^{(2)}\,
g_{\tilde{\varphi}_{2}\,l}^{(1)*}\,
\,\sum_{m}\,g_{\tilde{\varphi}_{1}\,m}^{(1)*}\,
g_{\tilde{\varphi}_{2}\,m}^{(2)})\,I^{(2)} \nonumber \\
&&+ c.c. \,,
\end{eqnarray}
\begin{eqnarray}
\label{int2}
\Gamma_{\tilde{\varphi}_{1}^{*},V}^{int(1)}& =& 
(\sum_{l}\,g_{\tilde{\varphi}_{1}\,l}^{(1)*}\,
g_{\tilde{\varphi}_{2}\,l}^{(1)}\,
\,\sum_{m}\,g_{\tilde{\varphi}_{1}\,m}^{(1)}\,
g_{\tilde{\varphi}_{2}\,m}^{(1)*})\,I^{(1)} \nonumber \\
&&+ (\sum_{l}\,g_{\tilde{\varphi}_{1}\,l}^{(2)*}\,
g_{\tilde{\varphi}_{2}\,l}^{(1)}\,
\,\sum_{m}\,g_{\tilde{\varphi}_{1}\,m}^{(1)}\,
g_{\tilde{\varphi}_{2}\,m}^{(2)*})\,I^{(2)} \nonumber \\
&&+ c.c. \,.
\end{eqnarray}
where $g_{\tilde{\varphi}_{1,2}\,l}^{(1,2)}$ is defined in Eq. (\ref{yuk})
and $I^{(1,2)}$ is an integral in which $\psi^{(Z)}_{1,2}$ propagates in the loop.
($I^{(1,2)}$ will be explicitely given below.) The sums are over 
all three flavors of leptons. Similarly, for the process
$\tilde{\varphi}_{1}^{(Z)} \rightarrow \bar{\psi}^{(Z)}_{2}+ l$,
\begin{eqnarray}
\label{int3}
\Gamma_{\tilde{\varphi}_{1},V}^{int(2)}& =& 
(\sum_{l}\,g_{\tilde{\varphi}_{1}\,l}^{(2)}\,
g_{\tilde{\varphi}_{2}\,l}^{(2)*}\,
\,\sum_{m}\,g_{\tilde{\varphi}_{1}\,m}^{(2)*}\,
g_{\tilde{\varphi}_{2}\,m}^{(2)})\,I^{(2)} \nonumber \\
&&+ (\sum_{l}\,g_{\tilde{\varphi}_{1}\,l}^{(1)}\,
g_{\tilde{\varphi}_{2}\,l}^{(2)*}\,
\,\sum_{m}\,g_{\tilde{\varphi}_{1}\,m}^{(2)*}\,
g_{\tilde{\varphi}_{2}\,m}^{(1)})\,I^{(1)} \nonumber \\
&&+ c.c. \,,
\end{eqnarray}
\begin{eqnarray}
\label{int4}
\Gamma_{\tilde{\varphi}_{1}^{*},V}^{int(2)}& =& 
(\sum_{l}\,g_{\tilde{\varphi}_{1}\,l}^{(2)*}\,
g_{\tilde{\varphi}_{2}\,l}^{(2)}\,
\,\sum_{m}\,g_{\tilde{\varphi}_{1}\,m}^{(2)}\,
g_{\tilde{\varphi}_{2}\,m}^{(2)*})\,I^{(2)} \nonumber \\
&&+ (\sum_{l}\,g_{\tilde{\varphi}_{1}\,l}^{(1)*}\,
g_{\tilde{\varphi}_{2}\,l}^{(2)}\,
\,\sum_{m}\,g_{\tilde{\varphi}_{1}\,m}^{(2)}\,
g_{\tilde{\varphi}_{2}\,m}^{(1)*})\,I^{(1)} \nonumber \\
&&+ c.c. \,.
\end{eqnarray}

It then follows that $\epsilon^{\tilde{\varphi}_{1}}_{1,2}$ which are
proportional to the difference between (\ref{int1}) and (\ref{int2}),
and  between (\ref{int3}) and (\ref{int4}) respectively, look as follows
\begin{eqnarray}
\label{epsilon1}
\epsilon^{\tilde{\varphi}_{1}}_{1,V} &\propto& Im\{\sum_{l} 
g_{\tilde{\varphi}_{1}\,l}^{(1)}\,
g_{\tilde{\varphi}_{2}\,l}^{(1)*}\,
\sum_{m}\,g_{\tilde{\varphi}_{1}\,m}^{(1)*}\,
g_{\tilde{\varphi}_{2}\,m}^{(1)}\} \,
Im\{I^{(1)}\} \nonumber \\
&&+ Im\{\sum_{l} g_{\tilde{\varphi}_{1}\,l}^{(2)}\,
g_{\tilde{\varphi}_{2}\,l}^{(1)*}\,
\sum_{m}\,g_{\tilde{\varphi}_{1}\,m}^{(1)*}\,
g_{\tilde{\varphi}_{2}\,m}^{(2)}\} \,
Im\{I^{(2)}\} \,, \nonumber \\
\end{eqnarray}
where the subscript $V$ denotes the contribution coming from
the interference of the tree-level and vertex-correction to
the decay widths. Notice however that $\sum_{l} 
g_{\tilde{\varphi}_{1}\,l}^{(1)}\,
g_{\tilde{\varphi}_{2}\,l}^{(1)*}\,
\sum_{m}\,g_{\tilde{\varphi}_{1}\,m}^{(1)*}\,
g_{\tilde{\varphi}_{2}\,m}^{(1)}$ is real and its imaginary part 
therefore vanishes. We are then left with 
\begin{eqnarray}
\label{epsilon1}
\epsilon^{\tilde{\varphi}_{1}}_{1,V} &\propto&
Im\{\sum_{l} g_{\tilde{\varphi}_{1}\,l}^{(2)}\,
g_{\tilde{\varphi}_{2}\,l}^{(1)*}\,
\sum_{m}\,g_{\tilde{\varphi}_{1}\,m}^{(1)*}\,
g_{\tilde{\varphi}_{2}\,m}^{(2)}\} \,
Im\{I^{(2)}\} \,. \nonumber \\
\end{eqnarray}
Similarly, with $\sum_{l} 
g_{\tilde{\varphi}_{1}\,l}^{(2)}\,
g_{\tilde{\varphi}_{2}\,l}^{(2)*}\,
\sum_{m}\,g_{\tilde{\varphi}_{2}\,m}^{(1)*}\,
g_{\tilde{\varphi}_{2}\,m}^{(2)}$ being real, we obtain
\begin{eqnarray}
\label{epsilon2}
\epsilon^{\tilde{\varphi}_{1}}_{2,V} &\propto&
Im\{\sum_{l} g_{\tilde{\varphi}_{1}\,l}^{(1)}\,
g_{\tilde{\varphi}_{2}\,l}^{(2)*}\,
\sum_{m}\,g_{\tilde{\varphi}_{1}\,m}^{(2)*}\,
g_{\tilde{\varphi}_{2}\,m}^{(1)}\} \,
Im\{I^{(1)}\} \,. \nonumber \\
\end{eqnarray}
Let us define
\be
\label{C}
C= \sum_{l} g_{\tilde{\varphi}_{1}\,l}^{(2)}\,
g_{\tilde{\varphi}_{2}\,l}^{(1)*}\,
\sum_{m}\,g_{\tilde{\varphi}_{1}\,m}^{(1)*}\,
g_{\tilde{\varphi}_{2}\,m}^{(2)} \,.
\ee
It is then easy to see that the coefficient of the right-hand-side
of (\ref{epsilon2}) is just $C^{*}$ so that $Im C^{*} = - Im C$. One can
then rewrite Eqs. (\ref{epsilon1}) and (\ref{epsilon2}) as
\be
\label{newepsilon1}
\epsilon^{\tilde{\varphi}_{1}}_{1,V} = (Im C)\,Im\{I^{(2)}\} \,, 
\ee
\be
\label{newepsilon2}
\epsilon^{\tilde{\varphi}_{1}}_{2,V} = -(Im C)\,Im\{I^{(1)}\} \,. 
\ee

In addition, the contribution to
$\epsilon^{\tilde{\varphi}_{1}}_{1,2}$ coming from the
tree-level-self-energy interference is found to be
\begin{eqnarray}
\label{interself1}
\epsilon^{\tilde{\varphi}_{1}}_{1,S} &\propto&
Im\{\sum_{m}\,g_{\tilde{\varphi}_{1}\,m}^{(1)*}\,
g_{\tilde{\varphi}_{2}\,m}^{(1)}\sum_{n}\,g_{\tilde{\varphi}_{1}\,m}^{(2)}\,
g_{\tilde{\varphi}_{2}\,m}^{(2)*}\} \nonumber \\
&& \times (\frac{m_{\tilde{\varphi}_1}^2}{
m_{\tilde{\varphi}_1}^2 -m_{\tilde{\varphi}_2}^2})\frac{1}{16\,\pi} \,,
\end{eqnarray}
\begin{eqnarray}
\label{interself2}
\epsilon^{\tilde{\varphi}_{1}}_{2,S} &\propto&
Im\{\sum_{m}\,g_{\tilde{\varphi}_{1}\,m}^{(1)}\,
g_{\tilde{\varphi}_{2}\,m}^{(1)*}\sum_{n}\,g_{\tilde{\varphi}_{1}\,m}^{(2)*}\,
g_{\tilde{\varphi}_{2}\,m}^{(2)}\} \nonumber \\
&& \times (\frac{m_{\tilde{\varphi}_1}^2}{
m_{\tilde{\varphi}_1}^2 -m_{\tilde{\varphi}_2}^2})\frac{1}{16\,\pi} \,.
\end{eqnarray}
However, as discussed in \cite{hung2}, $m_{\tilde{\varphi}_1}^2 \ll
m_{\tilde{\varphi}_2}^2$ (the ratio is approximately $\sim 10^{-26}$),
the contribution of $\epsilon^{\tilde{\varphi}_{1}}_{1,2,S}$ to
$\epsilon^{\tilde{\varphi}_{1}}_{1,2}$ is negligible and we will neglect
it from hereon. In what follows, we will make the identification
\be
\label{iden}
\epsilon_{1,2} = \epsilon^{\tilde{\varphi}_{1}}_{1,2,V} \,,
\ee
and the total asymmetry is defined as
\be
\label{eptot}
\epsilon_{tot} = \epsilon_{1} + \epsilon_{2} \,.
\ee
Therefore, we will focus below on constraints coming
from the vertex corrections which will be given explicitly below. 

Before presenting the results of the integrations, we write
down explicitely the loop integrals $I$ in the case 
$m_{\psi^{(Z)}_{1,2}}= 0$ and $I^{(1,2)}$ in the case $m_{\psi^{(Z)}_{1,2}} \neq 0$.
All SM lepton masses are neglected in the propagators.
They are:
\begin{eqnarray}
\label{massless}
I& =& \int \frac{d^4 l}{(2\pi)^4}\frac{1}{(l+k)^2}\frac{1}{(l+k^{'})^2} \\ \nonumber
&& + \int \frac{d^4 l}{(2\pi)^4}\frac{m_{\tilde{\varphi}_{2}}^2}
{[(l+k)^2][(l+k^{'})^2][l^2 - m_{\tilde{\varphi}_{2}}^2]}\,,
\end{eqnarray}
and
\begin{eqnarray}
\label{massive}
I^{(1,2)} &=& \int \frac{d^4 l}{(2\pi)^4}\frac{1}{(l+k)^2}
\frac{1}{(l+k^{'})^2 - m_{\psi^{(Z)}_{1,2}}^2} \\ \nonumber
&& + \int \frac{d^4 l}{(2\pi)^4}\frac{m_{\tilde{\varphi}_{2}}^2}
{[(l+k)^2][(l+k^{'})^2 -m_{\psi^{(Z)}_{1,2}}^2][l^2 - m_{\tilde{\varphi}_{2}}^2]}\,,
\end{eqnarray}
where $k$ and $k^{'}$ are the four-momenta of the external
$\psi^{(Z)}_{1,2}$ and SM lepton $l$ respectively.
(In (\ref{massive}), the numerator of the second term actually should
be $m_{\tilde{\varphi}_{2}}^2 + 4\, m_l\,m_{\psi^{(Z)}_{1,2}}
\sim m_{\tilde{\varphi}_{2}}^2$.)

To make our discussion as concise as possible, we first present the result for
the limit $m_{\psi^{(Z)}_{1,2}} \rightarrow 0$ to simply show that it
coincides with the well-known results. For this purpose, let us define
\be
\label{x}
x \equiv (\frac{m_{\tilde{\varphi}_{2}}}{m_{\tilde{\varphi}_{1}}})^{2} \,.
\ee
In the limit $m_{\psi^{(Z)}_{1,2}} \rightarrow 0$, one has
$Im\,I^{(1)}=Im\,I^{(2)}= Im\,I$ where
\be
\label{integral2}
Im\,I= \frac{1}{16\,\pi}(1-x\,\ln (1+ \frac{1}{x}))\,.
\ee
A few remarks are in order here in order to clarify the contrast
of the results of our model with those of ``standard scenarios''.
In the above result for $Im\, I$, the mass-independent part
$1/16\,\pi$ (first term on the right-hand side of (\ref{integral2})
comes from the absorptive part of the first integral in (\ref{massless}),
while the mass-dependent second term comes from the absorptive part
of the second integral in (\ref{massless}).
For $x \gg 1$, $(1-x\,\ln (1+ \frac{1}{x}))= (1-x(\frac{1}{x} -
\frac{1}{2\,x^2}+..) \sim \frac{1}{2\,x}$ and 
one obtains the familiar result, namely
$Im\,I \sim \frac{1}{32\,\pi \,x}$. The mass-independent term cancels
with the first term in the expansion of the logarithm.
At this point, one might
want to make contact with results that one obtains from an effective
theory after integrating out the heavy degree of freedom, namely
$\tilde{\varphi}_{2}$. In an effective 
theory where the heavy degree of freedom is integrated out so that
${\cal L} = {\cal L}_{tree} + (1/m_{\tilde{\varphi}_{2}}^2) 
{\cal L}_{dim 6}+...$, one can construct an 
equivalent one-loop vertex correction and obtains the suppression factor
$m_{\tilde{\varphi}_{1}}^2/m_{\tilde{\varphi}_{2}}^2$ directly without 
the aforementioned cancellation. In
that sense, the effective theory "misses" that mass-independent term
$1/ 16 \pi$ which should be there in the full one loop calculation
(used in most papers) but which gets cancelled for $x \gg 1$.

When $m_{\psi^{(Z)}_{1,2}}$ is not too different from $m_{\tilde{\varphi}_{1}}$,
there are some differences with the above ``massless'' results.
The first term of Eq. (\ref{massive}) gives an identical absorptive part to 
to that of the first term of Eq. (\ref{massless}), namely $(1/ 16 \pi)$.
The absorptive part of the second integral in (\ref{massive}) is now
proportional to $ln [(m_{\tilde{\varphi}_{2}}^2/m_{\tilde{\varphi}_{1}}^2 +1 
-m_{\psi^{(Z)}_{1,2}}^2/m_{\tilde{\varphi}_{1}}^2)/
(m_{\tilde{\varphi}_{2}}^2/m_{\tilde{\varphi}_{1}}^2 )]$.
The first
two terms inside the log are the well-known result for massless
fermions while the last term cannot be neglected in this case because
$m_{\psi^{(Z)}_{1,2}} = O(m_{\tilde{\varphi}_{1}})$, in contrast 
with previous scenarios of baryogenegis or leptogenesis. 
This gives rise to
an additional contribution in the vertex correction, namely
\be
\label{integral}
Im\,I^{(1,2)} = \frac{1}{16\,\pi}(1-x\,\ln (1+ \frac{1}{x} - \frac{1}{y_{1,2}})) \,,
\ee
where
\be
\label{y}
y_{1,2} \equiv (\frac{m_{\tilde{\varphi}_{2}}}{m_{\psi^{(Z)}_{1,2}}})^{2} \,.
\ee
For $x, y_{1,2} \gg 1$, one can again expand the logarithm term as
$1 -x ln(1+1/x - 1/y_{1,2}) = 1 -x(1/x -1/y_{1,2} -(1/2)(1/x - 1/y_{1,2})^2+..)
\sim x/y_{1,2} +(1/2)\,x\,(1/x - 1/y_{1,2})^2+..$. Again, the
mass-independent term cancels again $x(1/x)$ but now there remains an {\em extra}
term $x/y_{1,2}$ which would vanish if $m_{\psi^{(Z)}_{1,2}}= 0$. The second
term in the expansion is now a {\em subdominant} term equal to
$(1/2)(m_{\psi^{(Z)}_{1,2}}^2 - m_{\tilde{\varphi}_{1}}^2)^2/
(m_{\tilde{\varphi}_{1}}^2\,m_{\tilde{\varphi}_{2}}^2)$.
Taking into account the explicit definitions of $x$ and $y_{1,2}$, one can
expand Eq. \ref{integral} to find
\be
\label{integral2}
Im\,I^{(1,2)} = \frac{1}{16\,\pi}((\frac{m_{\psi^{(Z)}_{1,2}}}
{m_{\tilde{\varphi}_{1}}})^2 +...) \,,
\ee
where $...$ in Eq. \ref{integral2} denotes terms of order
$(m_{\tilde{\varphi}_{1}}/m_{\tilde{\varphi}_{2}})^2 \, < 10^{-26}$,
$(m_{\psi^{(Z)}_{1,2}}/m_{\tilde{\varphi}_{2}})^2 \, < 10^{-26}$ and higher
which are negligible compared with 
$(\frac{m_{\psi^{(Z)}_{1,2}}}{m_{\tilde{\varphi}_{1}}})^2 =x/y_{1,2}$. 
The dominant term in (\ref{integral2}) which comes from the
full one-loop calculation and which cannot
be neglected because $m_{\psi^{(Z)}_{1,2}} = O(m_{\tilde{\varphi}_{1}})$, 
{\em cannot} be seen from an effective theory.
This might appear to be "unexpected" because
in previous calculations, the decaying particle is much more
massive than any SM particles (either in SU(5) baryogenesis or
in standard scenarios of leptogenesis).
It is this quatity which will
determine the size of $Im\,I^{(1,2)}$ and hence that of the
asymmetry $\epsilon_{1,2}$. From (\ref{integral2}), one notices
that $Im\,I^{(1,2)}$ is {\em not} sensitive to the value
of the mass of $\tilde{\varphi}_{2}$ as long as
$m_{\tilde{\varphi}_{2}} \gg m_{\tilde{\varphi}_{1}}, m_{\psi^{(Z)}_{1,2}}$

From the above discussions, one notices the clear distinction between
our model for leptogenesis and the ``standard scenario'' involving
heavy Majorana neutrinos. Let us enumerate the differences. 
\bi

\item For the ``standard'' scenario,
SM particle masses are neglected compared with the heavy
Majorana neutrino masses in the one-loop computations
of the vertex and wave function corrections. As a result, the
lepton number asymmetry in these scenarios depend only on the ratios
of heavy Majorana neutrino masses, apart from the couplings and CP phase(s).

\item In our scenario, the decaying particle which gives rise to
the net lepton asymmetry is the messenger scalar field $\tilde{\varphi}_{1}$
whose mass is within the range of the electroweak scale while
the other messenger field $\tilde{\varphi}_{2}$ has a mass
of the order of a typical GUT scale. If these were the only
particles that one takes into account in the computation of the asymmetry,
the ratio of the mass of $\tilde{\varphi}_{1}$ to that of
$\tilde{\varphi}_{2}$ would negligibly small to play any role
in the asymmetry. The difference with the ``standard'' scenario lies
in the existence of the $SU(2)_Z$ fermions whose masses are not
too different from that of the decaying $\tilde{\varphi}_{1}$. This
gives, as a result, an asymmetry which mainly depends on the ratio
of the masses of the $SU(2)_Z$ fermions to that of $\tilde{\varphi}_{1}$
and which is no longer supressed by large mass ratios
such as $(m_{\tilde{\varphi}_{2}}/m_{\tilde{\varphi}_{1}})^2$
and $(m_{\tilde{\varphi}_{2}}/m_{\psi^{(Z)}_{1,2}})^2$. In fact,
the asymmetry  is {\em not} sensitive
to mass of the very heavy messenger field but depends instead on
the ratio of the masses of the two ``lighter'' particles.

\ei 

Let us now turn to the question of how big or how small $\epsilon_{tot}$
might be. In particular, it would be illuminating to see under what
conditions $\epsilon_{tot}$ {\em vanishes} so that we might learn
about the reasons why it does not vanish.
From the definitions (\ref{asymmetry1}, \ref{asymmetry2})
and the results (\ref{newepsilon1}, \ref{newepsilon2}), we
obtain
\begin{eqnarray}
\label{eptotal2}
\epsilon_{tot}& =& \epsilon_{1} + \epsilon_{2} \nonumber \\
&=& \frac{Im C}{\sum_{i}\,|g^{(1)}_{\tilde{\varphi}_{1}\,i}|^2}\,
Im\{I^{(2)}\} - \frac{Im C}{\sum_{i}\,|g^{(2)}_{\tilde{\varphi}_{1}\,i}|^2}\,
Im\{I^{(1)}\} \,. \nonumber \\
\end{eqnarray}
Using the results for $Im\{I^{(1,2)}\}$ (Eq. (\ref{integral2})), we can
rewrite (\ref{eptotal2}) as
\begin{eqnarray}
\label{eptotal3}
\epsilon_{tot}& =&\frac{1}{16\,\pi}\,(\frac{m_{\psi^{(Z)}_{2}}^2 -
m_{\psi^{(Z)}_{1}}^2}{m_{\tilde{\varphi}_{1}}^2})
\{\frac{Im C}{\sum_{i}\,|g^{(1)}_{\tilde{\varphi}_{1}\,i}|^2} \} \nonumber \\
&& + \frac{1}{16\,\pi}\,(\frac{m_{\psi^{(Z)}_{1}}^2}{m_{\tilde{\varphi}_{1}}^2})\,
(Im C)\,\{\frac{1}{\sum_{i}\,|g^{(1)}_{\tilde{\varphi}_{1}\,i}|^2}-
\frac{1}{\sum_{i}\,|g^{(2)}_{\tilde{\varphi}_{1}\,i}|^2} \}\,. \nonumber \\
\end{eqnarray}
A close examination of (\ref{eptotal3}) reveals some interesting
features of our model.
\bi

\item $\epsilon_{tot}$ vanishes if $Im C = 0$. This fact is self-evident
since, in order to obtain a non-vanishing asymmetry, one has to have 
CP-violating complex Yukawa couplings and hence $Im C \neq 0$ in our model.

\item $\epsilon_{tot}$ vanishes, even when $Im C \neq 0$, if we have {\em both}
\be
\label{equal1}
\frac{1}{\sum_{i}\,|g^{(1)}_{\tilde{\varphi}_{1}\,i}|^2} =
\frac{1}{\sum_{i}\,|g^{(2)}_{\tilde{\varphi}_{1}\,i}|^2} \,,
\ee 
and
\be
\label{equal2}
m_{\psi^{(Z)}_{1}} = m_{\psi^{(Z)}_{2}} \,.
\ee

\ei

Point \# 2 is quite interesting in its own right. Since, in the limit
(\ref{equal1}, \ref{equal2}), $\epsilon_{tot}=0$, this seems to suggest some 
kind of symmetry in the shadow sector involving $\psi^{(Z)}_{1,2}$. 
To see what this might be, let us again look at Eq. (\ref{eptotal3}). If we
put
\be
\label{g12}
|g^{(1)}_{\tilde{\varphi}_{1}\,i}| = |g^{(2)}_{\tilde{\varphi}_{1}\,i}| \,,
\ee 
in (\ref{eptotal3}), we obtain
\be
\label{eptotal4}
\epsilon_{tot}=\frac{1}{16\,\pi}\,(\frac{m_{\psi^{(Z)}_{2}}^2 -
m_{\psi^{(Z)}_{1}}^2}{m_{\tilde{\varphi}_{1}}^2})
\{\frac{Im C}{\sum_{i}\,|g^{(1)}_{\tilde{\varphi}_{1}\,i}|^2} \}  \,. 
\ee
The equality (\ref{g12}) suggests the existence of a {\em custodial symmetry}, 
at {\em tree-level} (where only the magnitudes of the couplings are involved),
in the shadow
sector which is explicitely {\em broken} by the mass difference between $\psi^{(Z)}_{1}$
and $\psi^{(Z)}_{2}$. Let us assume this custodial symmetry to be a global
$SU(2)$ and let us denote it by $SU(2)_{shadow}$. The shadow fermions then belongs
to a doublet of $SU(2)_{shadow}$: $(\psi^{(Z)}_{1}, \psi^{(Z)}_{2})$. The breaking
of this custodial symmetry by the shadow mass difference such that 
$\epsilon_{tot} \neq 0$ is reminescent of the breaking of the SM custodial
symmetry by mass differences among the up and down members of the $SU(2)_L$
doublet such that $\rho \neq 1$. The factor 
$(\frac{m_{\psi^{(Z)}_{2}}^2 -
m_{\psi^{(Z)}_{1}}^2}{m_{\tilde{\varphi}_{1}}^2})$
plays a similar role to the radiative correction to the $\rho$ parameter
coming from, e.g. a quark doublet, which is proportional to
$(\frac{m_{D}^2 - m_{U}^2}{m_{W}^2})$.

Let us come back to the statement that $\epsilon_{tot}=0$ when
$Im C = 0$ (Point \# 1). This can be achieved when
\be
\label{phase3}
g_{\tilde{\varphi}_{1}\,m}^{(1)}=g_{\tilde{\varphi}_{1}\,m}^{(2)}=
|g_{\tilde{\varphi}_{1}\,m}|
\exp (i \alpha_{m}) \,,
\ee 
\be
\label{phase4}
g_{\tilde{\varphi}_{2}\,m}^{(1)}=g_{\tilde{\varphi}_{2}\,m}^{(2)}=
|g_{\tilde{\varphi}_{2}\,m}|
\exp (i \beta_{m}) \,.
\ee 
With (\ref{phase3}, \ref{phase4}), one can easily see that $C$ is real
and hence $Im C = 0$. What (\ref{phase3}, \ref{phase4}) also imply is that
the shadow custodial symmetry $SU(2)_{shadow}$, when it is applied beyond
the tree-level to the Yukawa couplings, gives rise to $\epsilon_{tot}=0$
even if it is broken in the mass sector by having
$m_{\psi^{(Z)}_{1}} \neq m_{\psi^{(Z)}_{2}}$.

Last but not least, one might want to know whether or not we can put
$\tilde{\varphi}_{1}$ and $\tilde{\varphi}_{2}$ together into a doublet
of the shadow custodial symmetry. In what follows, we will see that
the constraint from the $K$ factor appears to rule out such a possibility.
In the last section on phenomenology, we will be pointing out that
from the constraint $0.04<K<1$ (\ref{lower}), one can deduce that
$8 \times 10^{-17} <\alpha_{\tilde{\varphi}_{1}} <2 \times 10^{-15}$.
The shadow custodial symmetry that includes 
$\tilde{\varphi}_{1}$ and $\tilde{\varphi}_{2}$
would then imply $\alpha_{\tilde{\varphi}_{1}}
= \alpha_{\tilde{\varphi}_{2}}$. As seen below, the asymmetry is
proportional to $\alpha_{\tilde{\varphi}_{2}}$ and the K factor constraint
would make it many orders of magnitude below the required value if
the shadow custodial symmetry also includes the messenger fields.

The above discussions suggest a deep connection between the possible
existence of a custodial symmetry in the shadow sector and the size
of the lepton number asymmetry $\epsilon_{tot}$: the breaking of that
custodial symmetry at loop levels gives rise to a {\em non-vanishing}
asymmetry. Our next step is to make an estimate for the asymmetry as
a function of various quantities which might have direct 
phenomenological implications such as the messenger mass and the shadow
fermion masses.

As mentioned above, the constraint we will use is $|\epsilon_{tot}| \sim 10^{-7}$.
Also for simplicity, we will use
the formula (\ref{eptotal4}) for $\epsilon_{tot}$. 
Let us first define
\be
\label{deltam2}
|\Delta m_{\psi^{(Z)}}^2| \equiv |m_{\psi^{(Z)}_{2}}^2 -
m_{\psi^{(Z)}_{1}}^2| \,.
\ee
\bi
\item As an example, let us take some numbers presented in Fig.(\ref{fig1}), namely
$m_{\tilde{\bm{\varphi}}_{1}^{(Z)}}
= 300\,GeV$, $m_{\psi^{(Z)}_{2}} = 100\; GeV$ and
$m_{\psi^{(Z)}_{1}}= 50\;GeV$. We then obtain
\be
\frac{1}{16\,\pi}\,(\frac{|\Delta m_{\psi^{(Z)}}^2|}
{m_{\tilde{\varphi}_{1}}^2}) \sim 1.7 \times 10^{-3}\,.
\ee
One can then obtain the following estimate
\be
\label{factor1}
\frac{|Im C|}{\sum_{i}\,|g^{(1)}_{\tilde{\varphi}_{1}\,i}|^2} \sim 10^{-4}\,.
\ee
For the sake of estimation, let us assume that 
$|g_{\tilde{\varphi}_{1}\,e}|\sim|g_{\tilde{\varphi}_{1}\,\mu}|\sim
|g_{\tilde{\varphi}_{1}\,\tau}|= |g_{\tilde{\varphi}_{1}}|$ and
$|g_{\tilde{\varphi}_{2}\,e}| \sim |g_{\tilde{\varphi}_{2}\,\mu}|
\sim |g_{\tilde{\varphi}_{2}\,\tau}|= |g_{\tilde{\varphi}_{2}}|$. One then
obtains
\be
\label{factor2}
\frac{|Im C|}{\sum_{i}\,|g^{(1)}_{\tilde{\varphi}_{1}\,i}|^2} \sim 
|g_{\tilde{\varphi}_{2}}|^{2} \sin \chi \sim 10^{-4}\,,
\ee
where $\sin \chi$ is a function of the various phases. Notice that (\ref{factor2})
is practically independent of the size of the Yukawa couplings of the lighter
decaying messenger field.

\item To obtain a rough estimate on the {\em lower} bound on the
light messenger field mass, we set 
$|g_{\tilde{\varphi}_{2}}|^{2} \sin \chi < 1$, giving the following bound,
for $m_{\psi^{(Z)}_{2}} = 100\; GeV$ and
$m_{\psi^{(Z)}_{1}}= 50\;GeV$,
\be
\label{bound1}
m_{\tilde{\bm{\varphi}}_{1}^{(Z)}} < 38 \, TeV \,.
\ee

\item A repeat of the above estimate with $m_{\psi^{(Z)}_{2}} = 100\; GeV$ and
$m_{\psi^{(Z)}_{1}}= 98\;GeV$, for example, yields 
\be
\label{bound2}
m_{\tilde{\bm{\varphi}}_{1}^{(Z)}} < 9 \, TeV \,.
\ee

\item Our last example is with $m_{\psi^{(Z)}_{2}} = 100\; GeV$,
$m_{\psi^{(Z)}_{1}}= 50\;GeV$ and
$|g_{\tilde{\varphi}_{2}}|^{2} \sin \chi \sim 10^{-3}$. This gives 
\be
\label{bound3}
m_{\tilde{\bm{\varphi}}_{1}^{(Z)}} < 1.2 \, TeV \,.
\ee

\ei

The next interesting question to ask is how small can
$|\Delta m_{\psi^{(Z)}}^2|$ be. First, the messenger field
cannot be too light since it has to decay while the sphaleron
process is still in thermal equilibrium as we have discussed above.
Using the constraint (\ref{lower}), let us for definiteness
set the minimum value for the mass of the messenger field
to be approximately $100\,GeV$. One now obtains a {\em lower bound}
on $|\Delta m_{\psi^{(Z)}}^2|$, namely
\be
\label{deltambound1}
|\Delta m_{\psi^{(Z)}}^2| > 0.05 \, GeV^2\,.
\ee
Using more ``reasonable'' values for 
$\frac{|Im C|}{\sum_{i}\,|g^{(1)}_{\tilde{\varphi}_{1}\,i}|^2}$, say $10^{-3}$,
and $m_{\tilde{\bm{\varphi}}_{1}^{(Z)}} \sim 300\, GeV$, one obtains
\be
\label{deltambound1}
|\Delta m_{\psi^{(Z)}}^2| > 452 \, GeV^2\,.
\ee
From the above estimates, one can infer that, unless there is a high
degree of degeneracy in the shadow fermion sector, the masses
$m_{\psi^{(Z)}_{i}}$ can be naturally in the range which is
suitable for them to be candidates for Cold Dark Matter (CDM), namely
$O(100\,GeV)$.

What the previous estimates show is that, for a given value of
the factor $|g_{\tilde{\varphi}_{2}}|^{2} \sin \chi$, the more degenerate
$\psi^{(Z)}_{1}$ and $\psi^{(Z)}_{2}$ are, the lower the
$\tilde{\bm{\varphi}}_{1}^{(Z)}$ mass should be in order to get
a reasonable value for the asymmetry. Turning the argument around,
one infers that the shadow fermions cannot be too degenerate
and that their masses can naturally be in the favored range
to be CDM candidates.

The above upper bounds on the light messenger mass can be lowered if we
allow for the factor $K$ (see Eq. \ref{K}) to be greater
than unity. Because of the dilution factor $1/K$, $\epsilon_{tot}$
changes to $\sim -10^{-7}\,K$. For example, the bound (\ref{bound2})
is lowered to $2.8\, TeV$ and $890\, GeV$ for $K=10, 100$ respectively.


The examples given above are far from being exhaustive and are simply meant
to be illustrative of the deep connection, in our model, between the SM
lepton asymmetry, which is eventually transmogrified into a baryon
asymmetry, and the mass of the messenger scalar field $\tilde{\varphi}_{1}$
responsible for this asymmetry.
They point to the fact that $\tilde{\varphi}_{1}$ could  be relatively
``light'' and has thus a ``chance'' to be found if
it exists. In addition, it was also shown that a deep connection
exists between the asymmetry and the breaking of the shadow custodial
symmetry discussed above.

As we have mentioned above, a complete treatment of the asymmetry linking
$\epsilon_{tot}$ to the actual SM lepton number asymmetry, and eventually
to the baryon asymmetry, requires one to solve the Boltzman equation taking
into account various factors such as decays and inverse decays, etc... This
however will not significantly change the various bounds derived above.
A more detailed study will be presented elsewhere.

Last but not least, we would like to remark that the lepton flavour effects
encountered in see-saw leptogenesis scenarios
\cite{flavour} do not affect our model since the decays here are very
out-of-equilibrium and the single-flavour analysis is justified.

\section{Phenomenological consequences of the ``lepton number progenitor'' 
$\tilde{\varphi}_{1}^{(Z)}$
with mass $\alt 1\,TeV$}

Since this section is slightly out of the main topic of the paper, it
will be very short and the details will be presented elsewhere.  
The main purpose for including it here is to show 
that there are consequences of our proposed SM leptogenesis scenario
that can be tested experimentally in a not-too-distant future at the
LHC. The progenitor for the aforementioned lepton asymmetry can 
possibly be found and identified experimentally! 
As we had mentioned in Section (\ref{review}), we require
$SU(2)_Z$ to be confining and, as the result, the messenger fields
which carry both $SU(2)_Z$ and electroweak quantum numbers cannot have
a vacuum expectation value. In the kinetic terms for the messenger fields,
and in particular for $\tilde{\varphi}_{1}^{(Z)}$, one is interested in the
following interactions: $W^{+}\,W^{-}\,(\tilde{\varphi}_{1}^{(Z),0*}
\tilde{\varphi}_{1}^{(Z),0}+\tilde{\varphi}_{1}^{(Z),+}\,
\tilde{\varphi}_{1}^{(Z),-})$
and $Z\,Z\,(\tilde{\varphi}_{1}^{(Z),0*}
\tilde{\varphi}_{1}^{(Z),0}+\tilde{\varphi}_{1}^{(Z),+}\,
\tilde{\varphi}_{1}^{(Z),-})$.
These interactions will provide the dominant weak boson fusion (WBF)
production mechanism for a pair of $\tilde{\varphi}_{1}^{(Z)}$. A rough
expectation for the production cross section for $\tilde{\varphi}_{1}^{(Z)}$
with a mass around $300\,GeV$ is around $1\,pb$ and $0.1\,pb$
for a mass around $500\,GeV$. The decay
$\tilde{\varphi}_{1}^{(Z),0} \rightarrow \bar{\psi}^{(Z)}_{1,2}+ l^{0}_{i}$
is practically unobservable while
$\tilde{\varphi}_{1}^{(Z),-} \rightarrow \bar{\psi}^{(Z)}_{1,2}+ l^{-}_{i}$
and $\tilde{\varphi}_{1}^{(Z),+} \rightarrow \psi^{(Z)}_{1,2}+ l^{+}_{i}$
will have charged SM leptons with unconventional geometry, perfectly
distinguishable from the decay of a $600\,GeV$ SM Higgs boson.

For the decays of $\tilde{\varphi}_{1}^{(Z),\pm}$, one might want to have
a rough idea on the length of the charged tracks before the decays occur.
For definiteness, let us take $m_{\tilde{\varphi}_{1}} = 500\,GeV$
as an example. From the definition of $K$ (\ref{K}) and from the
requirement $0.04<K<1$ (\ref{lower}), one can deduce that
$8 \times 10^{-17} <\alpha_{\tilde{\varphi}_{1}} <2 \times 10^{-15}$.
Since $\Gamma_{\tilde{\varphi}_{1}}
\sim \alpha_{\tilde{\varphi}_{1}} m_{\tilde{\varphi}_{1}}$, the decay
length is roughly $0.02\,cm<l_{\tilde{\varphi}_{1}}<0.5\,cm$. This
decay length falls within the range of the radial
region of a typical silicon detector at CMS 
and ATLAS ($40\,cm$ and $60\,cm$ respectively). Notice that
when $K>1$ implying a larger $\alpha_{\tilde{\varphi}_{1}}$, the
decay length is even smaller than the previous upper bound, again
well within reach of the aforementioned silicon detectors.
It is also conceivable that, if these decays were to be observed,
one might be able to measure the CP violating phases in 
Eq. (\ref{yuk}) and, as a consequence, the size of the SM
lepton number asymmetry needed in this leptogenesis scenario.
It is interesting to note that our scenario allows for
a {\em direct} search of the {\em progenitor} of the
lepton, and hence baryon, asymmetry at future colliders.

The detection of $\bar{\psi}^{(Z)}_{1,2}$ would fall into the domain
of Dark Matter search since it is electrically neutral and 
interacts very weakly with normal matter \cite{hung2}. However,
it can also be indirectly ``observed'' as missing energy in the
decay of the messenger field.

One might also ask whether or not the process $\mu \rightarrow
e \gamma$ can be affected by the couplings discussed in this paper.
It can occur at one loop with $\psi^{(Z)}_{1,2}$ and 
$\tilde{\varphi}_{1}$ or $\tilde{\varphi}_{2}$ propagating in the loop.
First there is no enhancement factor encountered in $Im\,I^{(1,2)}$.
Second the rate is {\em negligibly small} because of constraints
such as $8 \times 10^{-17} <\alpha_{\tilde{\varphi}_{1}} <2 \times 10^{-15}$
or in the second case because of the suppression factor
$(\frac{m_{\psi^{(Z)}_{1,2}}}{m_{\tilde{\varphi}_{2}}})^{2}$.

Finally, we wish to point out that our shadow fermions $\psi^{(Z)}_{1,2}$
do not acquire a millicharge which would have happen if
$SU(2)_Z$ is broken down to $U(1)$ which could mix through
vacuum polarization with $U(1)_{em}$ \cite{holdom}. In our model $SU(2)_Z$
is unbroken and this is the reason why it grows strong
at a very low scale $\sim 10^{-3}\,eV$.

\begin{acknowledgments}
This work is supported in parts by the US Department
of Energy under grant No. DE-A505-89ER40518. 
I wish to thank Lia Pancheri, Gino Isidori and
the Spring Institute for the hospitality in the
Theory Group at LNF, Frascati, where part of this work
was carried out. I also wish to thank Manny Paschos
for useful comments concerning the first draft.
\end{acknowledgments}

\begin{figure}
\includegraphics[angle=-90,width=8cm]{alvst_100.epsi}
\includegraphics[angle=-90,width=8cm]{alinvst_100.epsi} 
\caption{\label{fig2}$\alpha_Z(E)$ and $\alpha_Z^{-1}(E)$ versus
$t=\ln (E/\Lambda_Z)$ for $m_{\tilde{\bm{\varphi}}_{1}^{(Z)}}
= 300\,GeV$, $m_{\psi^{(Z)}_{2}} = 200\; GeV$ and
$m_{\psi^{(Z)}_{1}}= 100\;GeV$. Here $\Lambda_Z = 3 \times 10^{-3}\,eV$.}
\end{figure}
\begin{figure}
\includegraphics[angle=0,width=4cm]{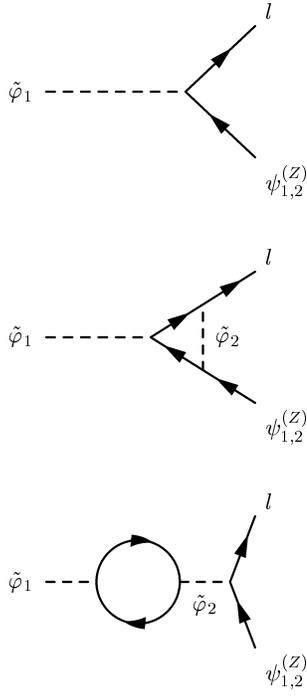}
\caption{\label{fig3}. The decay
$\tilde{\varphi}_{1}^{(Z)} \rightarrow \bar{\psi}^{(Z)}_{1,2}+ l$ at tree-level
and at one loop}
\end{figure}
\end{document}